# ANALYSIS OF THE BENDING BEHAVIOUR OF FLAX BASED REINFORCEMENTS USED IN SHAPE FORMING

A.Bassoumi[1]*, P.Ouagne[1], J. Gillibert[1], G. Hivet[1]

[1]PRISME Laboratory, F2ME Research Team, University of Orleans, France

*corresponding author: amal.bassoumi@univ-orleans.fr

## ABSTRACT

*The bending behaviour of woven perform was investigated in order to better understand the formation of some defects during sheet forming such as wrinkling and tow buckling. The fabric composition considering hybrid and pure flax fabrics as well as some test conditions like relative humidity were examined. On the one hand, the results show a drop of the bending stiffness with flax/PLA commingled fabric. On the other hand, the study points out that moisture enhances the bending rigidity especially in the case of pure flax fabric. However, an excess of humidity, for instance 100% relative humidity, leads to an opposite effect.*

## INTRODUCTION

Composite materials are considered as an essential component of the tomorrow's industry. Possessing good mechanical properties, they also reduce energy consumption thanks to a gain of weight. Besides, in a perspective of sustainable development, industry is attempting to replace existing ones by environmentally friendly composites. Disposal problems and the need to reduce the dependence on petroleum fuels, lead to an increasing interest for natural fibres based composites [1] and particularly the flax based ones offer a good potential.

In addition to tensile and shear phenomena widely considered for carbon or glass reinforcements, the bending behaviour was generally neglected when considering the sheet forming simulation of highly non-developable parts [2]. For structural parts, the properties out of plan cannot be inferred from those in plan. The flexural behaviour of the reinforcement depends on yarns geometry, their mechanical properties and contact conditions. Therefore, we are dealing with a multi-scale problem [3].

The bending rigidity is a critical parameter regarding the out of plane deformations at both macroscopic and mesoscopic scale such as wrinkles formation and tow buckling. Some studies on process simulation showed the importance of this rigidity for either dry woven reinforcements or prepregs.

For a long time, the bending behaviour of textile clothing has been considered. In this context, Ghosh et al presented a critical review of various models with different yarn behaviour assumptions [4, 5]. However, the relationship between fabric and yarn bending rigidities is highly complex. On the contrary, there are few research works regarding composite reinforcements: Bilbao et al [2], Yu et al [6] and Lomov et al [7].

The aim of this paper is to study the influence of the fabric composition as well as test conditions on the bending stiffness of woven perform. In fact, two compositions have been considered: hybrid and pure flax fabrics. The effect of environmental conditions is evaluated regarding various relative humidities.

## MATERIALS AND METHODS

### 1. MATERIALS

The dry reinforcements considered in this study are 4x4 hopsack fabrics (Figure 1). The first reference is pure flax based fabric with an areal weight of 494 g/m$^2$ and a thickness of 1.29 mm. The second reference is 40% flax/PLA commingled hopsack with an areal weight of 518 g/m$^2$ and a thickness of 1.38 mm.

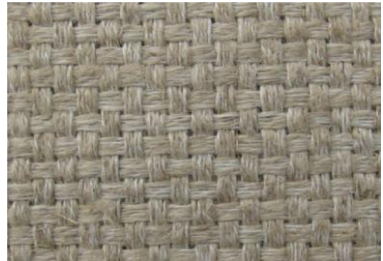

*Figure 1 The 4x4 hopsack fabric*

### 2. BENDING TEST ON FABRICS

Our experimental work was carried out using a bending test based on the cantilever's principle, in which the fabric bends under its own weight. The improved device allows various loading cases related to different bending lengths [2, 3].

The bending device used includes two modules: one mechanical and one optical. The first consists of two parts, one fixed (in the form of plate) and the other mobile composed of a set of removable strips rotatable around an axis coplanar with the fixed part. At the other end, the strips rely on a slide device for releasing them successively. To avoid friction between them, the strips are spaced by 2 mm using washers. After putting the sample on the horizontal assembly, a translucent blank holder plate is laid on it and clamped to the fixed plate which defines an embedded bond during bending.

The device allows at a growing number of released strips to increase the length of overhang. At a given length, the slide movement is stopped and an optical module comprising a CCD camera is used to capture the bending profile. After the test, a step of digital image processing is performed to extract by Framing-Filtering-Binarization the reference line. Then, using a calibration picture taken after the camera setting and knowing the real size of the image pixel, the scaling factor between the image space and the real space is determined. This information will be used to make the conversion pixel-mm.

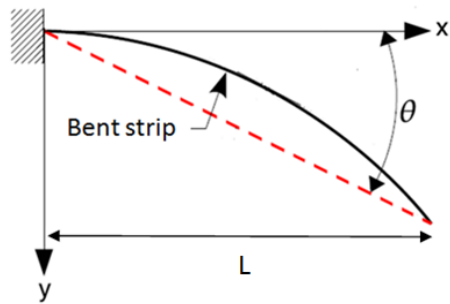

*Figure 2 The bending test for fabric: For a given bending length L, the rigidity depends on θ angle.*

Then, by a direct method, the curvature C and the bending moment M can be evaluated at each point of the profile using the digital processing. Therefore, the M-C relationship can be estimated. However, as we are looking to a comparative study of the influence the reinforcement parameters, this paper will focus on the pierce rigidity.

The processed images can be used to determine the Peirce model [8, 9] and thus the bending stiffness $G$. For this, the coordinates of the embedding bond and the free end points should be specified to calculate the $\theta$ angle (see Figure 2). For a fabric in cantilever configuration under gravity, Peirce defined the ratio $c^3$:

$$c^3 = \frac{G}{Ad\ w\ g}$$

with $Ad$ and $w$ are respectively the areal weight and the width of the fabric; $g$ the earth's standard acceleration due to gravity. $c$ has the unit of a length. It is named the bending length by Peirce.

Then, he suggested the following approximation:

$$c^3 = \frac{L^3}{8} \frac{\cos(\frac{\theta}{2})}{\tan \theta}$$

with $L$ and $\theta$ are respectively the length of the fabric bending part and the angle between the chord and the horizontal axis.

Thus, the stiffness goes to:

$$G = \frac{L^3}{8} \frac{\cos(\frac{\theta}{2})}{\tan \theta} Ad\ w\ g$$

During the test, the sample is about 30 mm width and 300 mm length. Nonetheless, due to a twist effect in the bending fabric, two profiles (top and bottom) could be considered (see Figure 3). The following study focuses on the top profile. Its rigidity is denoted by $Gs$.

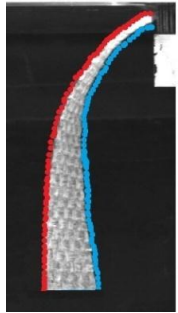

*Figure 3 Top and bottom profiles*

### 3. BENDING TEST IN VARIOUS RELATIVE HUMIDITY

As the materials studied in this work are based on natural fibres, the degree of moisture absorption may influence their mechanical properties and particularly the bending stiffness. The test samples are therefore prepared and submitted to different relative humidity $RH$ environments. For the dried fabrics, the following protocol was used: the samples were dried for 24 h at 110 °C before being tested inside the oven where RH is about 14 %. Besides, fabrics conditioned at about 100 % of humidity (RH=99.9%) were also tested. Finally, three intermediates humidities were examined (RH= 33, 55 and 86 %). In this context, the samples were put in climatic enclosures airtight with a solution in the bottom and a sample holder. For the solution, we use water for 100 % RH and aqueous saturated salt solutions for 33, 55 and 86 % RH. During the tests the room temperature was between 20 and 23 °C. According to standard NF EN ISO 483 2006-01 [10], depending on the chemical salt used and the room temperature, an environment with controlled humidity can be created (Table 1). After being conditioned for 24 h, the fabrics were tested under ambient conditions (RH=63 % on average). The time taken for test including the placement of the sample is about 5 min.

*Table 1 Relative humidities according to the salt used*

| Salt | RH for 20 °C |
|---|---|
| Magnesium chloride: Mg $Cl_2$ | 33 % |
| Calcium nitrate: Ca $(NO_3)_2$ | 55 % |
| Potassium chloride: KCl | 86 % |

We point out that the references were placed in different atmospheres (oven or climatic enclosures) for 24 h. In fact, the evolution in time of the references weight in each atmosphere was carried out. The figure 4 shows the evolution of the moisture content *MC* of the two references at intermediates humidities. Such study allowed us to detect the equilibrium and the time needed to weight stabilisation.

$$MC\ \% = \frac{w - w0}{w0}\ x\ 100$$

with *w* is the weight ; *w0* is the dry weight (the weight at time 0).

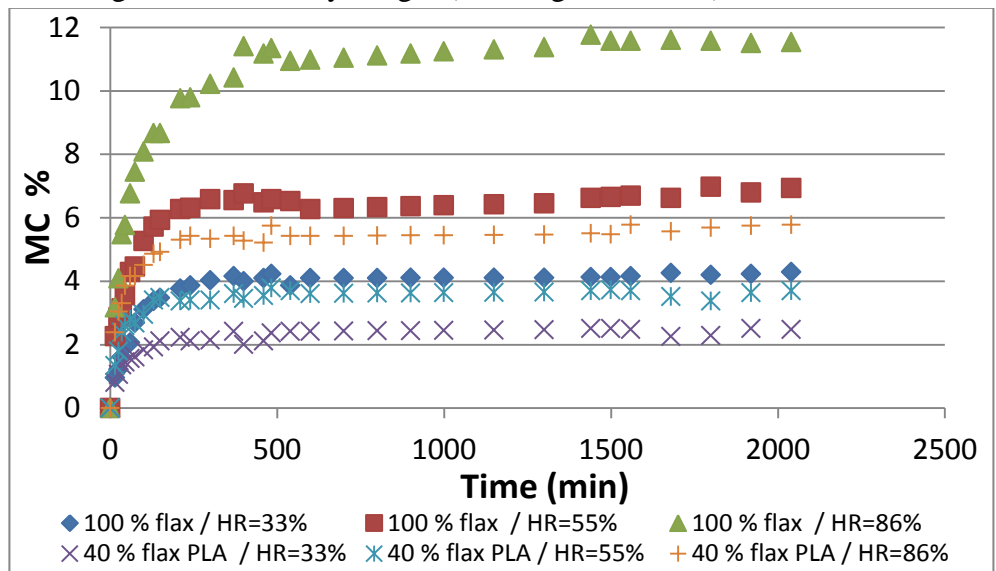

*Figure 4 The evolution in time of the moisture content at different relative humidities*

### RESULTS AND INTERPRETATION

#### 1. INFLUENCE OF THE FABRIC COMPOSITION

The fabric composition, considering hybrid and pure flax fabrics, was examined. Each time, five samples from the same roll were tested in the weft direction. Figure 5 represents the results of the bending rigidity of the 100% flax hopsack. It shows the variation of the rigidity Gs with bending length L. This can be used to check if the behaviour is linear elastic. In this case, the rigidity should be independent of the bending length [2].

The variation of the Gs amplitude ΔGs is about 79.68 % of the average which illustrates a non linear elastic behaviour with

$$\Delta Gs = \frac{Gs\max - Gs\min}{Gs\ average}$$

Assuming the general behaviour of Gs, it can be noted that the apparent stiffness is almost constant in the range of 100 to 180 mm

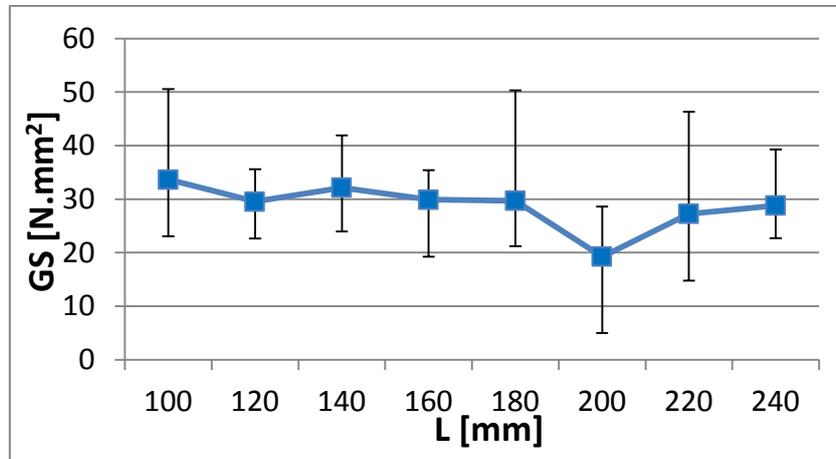

*Figure 5 The evolution of the bending stiffness Gs as a function of the bending length L for Hopsack 4x4 100% flax at ambient (RH=67.2% on average)*

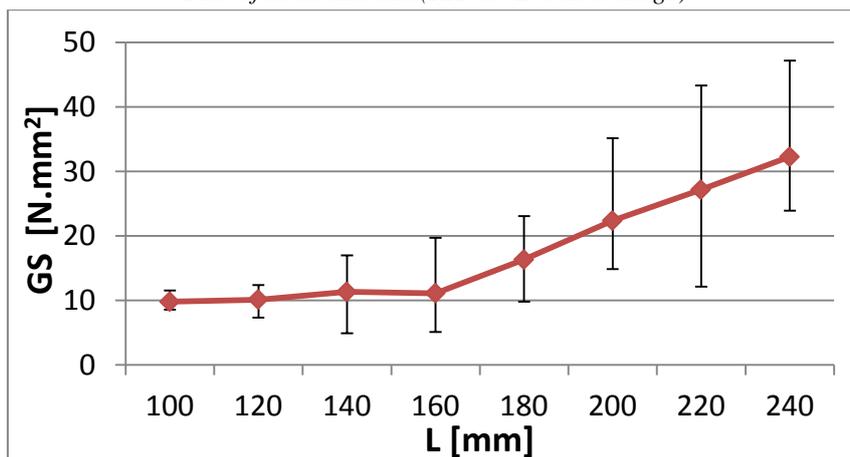

*Figure 6 The evolution of the bending stiffness Gs as a function of the bending length L for Hopsack 4x4 40% flax/PLA at ambient (RH=51.3% on average)*

In the case of 40% Flax/PLA, the variation of the Gs amplitude ΔGs is about 95.41 % of the average which illustrates a non linear elastic behaviour as well (Figure 6). Nevertheless, the apparent stiffness is almost constant in the range of 100 to 160 mm. Considering that the bending stiffness is representative of the general behaviour of the material between 100 and 160 mm, we observed the effect of the composition in this range.

For flax/PLA commingled hopsack a significant loss of stiffness (66% on average) is observed in comparison to the 100% flax fabric. The rigidity goes from 31 N.mm$^2$ with pure flax to 10 N.mm$^2$ with hybrid flax. Thus, with 60% of PLA the fabric becomes more flexible. Such result is expected in the case of homogeneous isotropic linear elastic material because the vertical displacement is inversely proportional to the young's modulus. Actually, the young's modulus of PLA, which is about 3 ± 0.15 GPa [11], is significantly lower than the young's modulus of flax which is about 58 ± 15 GPa [12].

Composite reinforcement shaping is a necessary step for process such as RTM and prepregs. In RTM, the matrix will be injected after forming such is the case of the pure flax fabric. For thermoplastic composite (CFRTP), a hot sheet forming of commingled fabric such as flax/PLA could be used. After the hot forming, thermoplastic fibres will play the role of the matrix. By reducing the bending rigidity, the adding of PLA can affect significantly the fabric acceptable curvature. According to this study, the hybrid fabric is more likely to form deflects such as wrinkles [13]. In this context, we cannot make the same parts with RTM and sheet forming unless to enhance the stiffness of the flax/PLA reinforcement and one perspective is to act on the humidity effect.

## 2. INFLUENCE OF THE RELATIVE HUMIDITY

Due to the hydrophilic nature of flax fibres [14], the evolution of bending properties with humidity was investigated. As it was observed before the apparent stiffness is almost constant between 100 and 160 mm. Thus, we will focus on a fixed bending length L=130 mm. Five relative humidities were examined: 14 %, 33 %, 55 %, 86 % and 100 % (Figure 7).

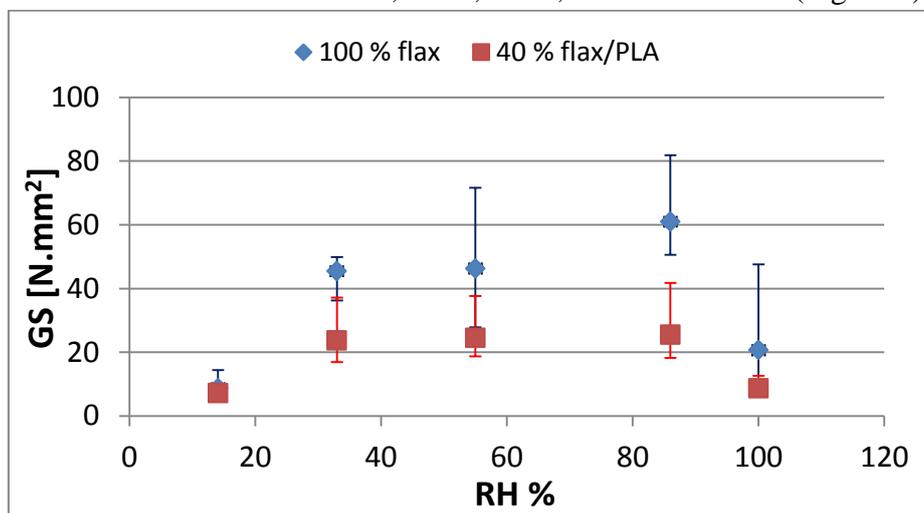

*Figure 7 Effect of the relative humidity RH on the bending stiffness Gs*

The graph shows that the rigidity increases as a function of a growing relative humidity until 86 % RH before dropping at 100 % RH. The water uptake is then advantageous for the rigidity up to a maximum relative humidity. The excess leads to an opposite effect. Up to 86 % RH, the rigidity growth could be explained by the plasticising effect of water possibly due to the presence of ''free water''. However, the excess of moisture increases the amount of ''bonded water'' in spite of free water [15].

Indeed, fabrics made from plant fibres are highly hygroscopic because they can host water in two forms: a free form which penetrates the cellulose network and is housed in capillaries and the spaces between fibrils; and a bound form attached to the cellulose molecules by chemical bonds. Polysaccharides are very hydrophilic [16]. Thanks to their hydroxyl group, they can host water molecules by Van der Waals and hydrogen bonds. In addition, the water molecules separate cellulose molecules. Hence, cellulose molecules are able to move more freely. The cellulose becomes more flexible and the insertion of water molecules in the hydrophilic network causes the swelling of the structure. The swelling can degrade the mechanical properties of the material by creating micro-cracks which can be confirmed by a SEM observation of the samples conditioned at 100% [15]. If a relative humidity of 100% results in a drop in the mechanical properties, a total water immersion can also cause a loss of stiffness [11, 17].

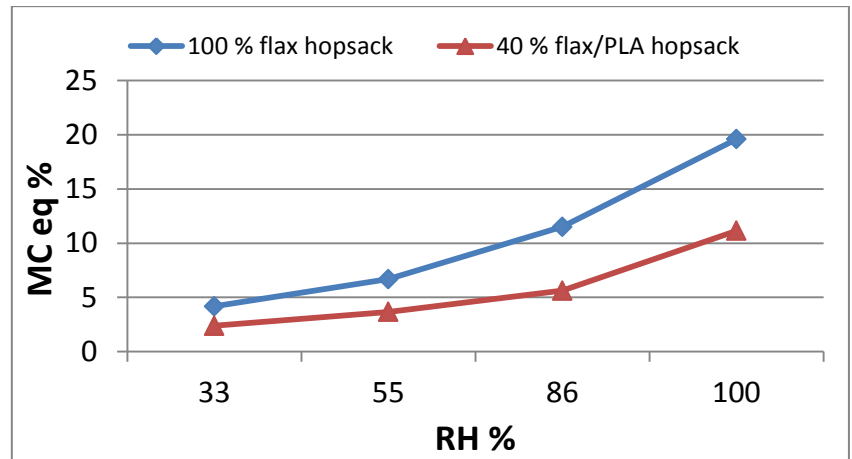

*Figure 8. The evolution of the moisture content at the equilibrium with the relative humidity*

As the sample conditioned at about 55% reveals the behaviour at ambient room conditions, moisture uptake can be disastrous (eg 100%) or advantageous (eg 86%). However, drying causes a loss of stiffness at either 14% or 33% RH. This is possibly due to the embrittlement of the fibre constituents after the removal of the plasticizer which is water [16, 18]. In addition, the water removal can cause mechanical stress on the scale of the fibre due to changes in behaviour of the fibre components (cellulose, pectin, hemicellulose) [19]. Moreover, the drying effect can be explained by the formation of hydrogen bonds between the cellulose surfaces limiting interaction between the fibrils [20, 21].

Previously, a gap of about 66 % in bending stiffness was noted between the two references under ambient conditions. Tested for various relative humidities, the 40% flax/PLA hopsack remains advanced by the pure flax fabric. In order to explain the drop in rigidity, the variation of samples water uptake with the air relative humidity was examined (Figure 8). Actually, at a given relative humidity, the moisture content differs from one reference to another. This is probably due to the difference in polar properties between PLA and cellulose. PLA is more hydrophobic than flax which is generally the case of the matrix in composite material [14]. And a previous study has point out that the PLA moisture content is about 2 % in the case of total immersion [11].

## CONCLUSION

The 40% flax/PLA reinforcement is more likely to generate defaults such as wrinkles during the sheet forming process because of a lower bending stiffness compared to the pure flax reinforcement. This can be explained by the PLA young's modulus which is lower than flax modulus as well as a less sensitivity to humidity.

This paper clearly shows that moisture enhances the bending rigidity especially in the case of 100% flax fabric. However, the study points out that the samples rigidity drops after conditioning in 100% relative humidity. This is probably due to fibre damage after an excessive water uptake. On the other hand, drying results in significant rigidity fall. The water removal embrittles the flax constituents and influence interfacial bonding between cellulose surfaces.

All things considered, we conclude that on the one hand, the environmental conditions during the process should be controlled. And on the other hand, acting only on the humidity, we cannot significantly improve the stiffness of flax/PLA reinforcement.